\pdfoutput=1

\documentclass[reprint,aps,pra,amsfonts,amsmath,amssymb,superscriptaddress]{revtex4-1}

\usepackage[utf8]{inputenc}
\usepackage{graphicx}
\usepackage[per-mode = symbol]{siunitx}
\usepackage{amssymb}
\usepackage{esint}
\usepackage{todonotes}
\usepackage{multirow}
\usepackage{vphys}
\usepackage{booktabs}

\AtBeginDocument{
\heavyrulewidth=.08em
\lightrulewidth=.05em
\cmidrulewidth=.03em
\belowrulesep=.65ex
\belowbottomsep=0pt
\aboverulesep=.4ex
\abovetopsep=0pt
\cmidrulesep=\doublerulesep
\cmidrulekern=.5em
\defaultaddspace=.5em
}

\makeatletter
\g@addto@macro{\UrlBreaks}{\UrlOrds}
\makeatother

\begin{document}


\title{Efficiently parallelized modeling of tightly focused, large bandwidth laser pulses}
\author{Joey Dumont}
\email{Corresponding author: joey.dumont@gmail.com}
\affiliation{Universit\'{e} du Qu\'{e}bec, INRS-\'{E}nergie, Mat\'{e}riaux et T\'{e}l\'{e}communications, Varennes, Qu\'{e}bec, Canada, J3X 1S2}

\author{François Fillion-Gourdeau}
\affiliation{Universit\'{e} du Qu\'{e}bec, INRS-\'{E}nergie, Mat\'{e}riaux et T\'{e}l\'{e}communications, Varennes, Qu\'{e}bec, Canada, J3X 1S2}
\affiliation{Institute for Quantum Computing, University of Waterloo, Waterloo, Ontario, Canada, N2L 3G1}

\author{Catherine Lefebvre}
\affiliation{Universit\'{e} du Qu\'{e}bec, INRS-\'{E}nergie, Mat\'{e}riaux et T\'{e}l\'{e}communications, Varennes, Qu\'{e}bec, Canada, J3X 1S2}
\affiliation{Institute for Quantum Computing, University of Waterloo, Waterloo, Ontario, Canada, N2L 3G1}

\author{Denis Gagnon}
\affiliation{Universit\'{e} du Qu\'{e}bec, INRS-\'{E}nergie, Mat\'{e}riaux et T\'{e}l\'{e}communications, Varennes, Qu\'{e}bec, Canada, J3X 1S2}
\affiliation{Institute for Quantum Computing, University of Waterloo, Waterloo, Ontario, Canada, N2L 3G1}

\author{Steve MacLean}
\email{steve.maclean@emt.inrs.ca}
\affiliation{Universit\'{e} du Qu\'{e}bec, INRS-\'{E}nergie, Mat\'{e}riaux et T\'{e}l\'{e}communications, Varennes, Qu\'{e}bec, Canada, J3X 1S2}
\affiliation{Institute for Quantum Computing, University of Waterloo, Waterloo, Ontario, Canada, N2L 3G1}

\date{\today}

\begin{abstract}
The Stratton-Chu integral representation of electromagnetic fields is used to study
the spatio-temporal properties of large bandwidth laser pulses focused by high numerical aperture mirrors.
We review the formal aspects of the derivation of diffraction integrals from the
Stratton-Chu re\-pre\-sen\-ta\-tion and discuss the use of the Hadamard finite part in the
derivation of the physical optics approximation. By analyzing the formulation
we show that, for the specific case of a parabolic mirror, the integrands
involved in the description of the reflected field near the focal spot do not possess
the strong oscillations characteristic of diffraction integrals.
Consequently, the integrals can be evaluated with simple and efficient quadrature methods
rather than with specialized, more costly approaches.
We report on the development of an efficiently parallelized algorithm that
evaluates the Stratton-Chu diffraction integrals for incident fields of arbitrary
temporal and spatial dependence. This method has the advantage that its input is
the unfocused field coming from the laser chain, which is experimentally known with
high accuracy.
We  use our method to show that the reflection of a linearly polarized Gaussian
beam of femtosecond duration off a high numerical aperture parabolic mirror
induces ellipticity in the dominant field components and generates strong longitudinal
components. We also estimate that future high-power laser facilities may reach
intensities of $10^{24}\,\si{\watt\per\cm\squared}$.
\end{abstract}

\maketitle

\section{Introduction}\label{sec:intro}

Recent proposals for high-power laser infrastructures such as ELI and APOLLON \cite{PAP2013,DAN2015}
have opened the possibility of an experimental detection of strong-field quantum electrodynamics (SF-QED)
effects including vacuum polarization, Breit-Wheeler pair production,
nonlinear Compton scattering and Schwinger pair production \cite{FED2009,MON2011,DIP2012,TIT2013,FIL2015a}.
The high intensities required to observe these effects can be approached by
using a temporally compressed high-power laser in conjunction with a tight
focusing scheme \cite{MOU2006,MOU2015}.

In the tight focusing regime, the spatial extension of the field is on the same order of magnitude as
its wavelength and, as a consequence, paraxial fields fail to properly model the
spatio-temporal properties of the field
at the focal spot \cite{LAX1975,SAL2006}. This approach generally considers an expansion
valid only for small diffraction angles $\epsilon\ll1$, where $\epsilon=\lambda/\pi w_0$
is the ratio of the wavelength of the beam  $\lambda$ to its transverse width $w_0$.
Tightly focused beams are characterized by $\epsilon\gg1$ and, in addition,
any realistic analysis in this regime must retain the vector nature of the fields.
Moreover, to properly represent the experimental reality,
the effect of the reflection on the mirror should be considered, requiring a formulation
that properly accounts for the boundary conditions on the reflecting surface.

Methods that consider the vector character of the fields have been used before,
starting with Ignatowski's study of Maxwell's equations
in parabolic cylindrical coordinates \cite{IGN1907,IGN1908}. Later, Kottler
\cite{KOT1923a,KOT1923b} formulated the reflection problem as an integral equation
using Green's function techniques. Stratton and Chu \cite{STR1939} then
generalized Kottler's result and streamlined its derivation.
Richards and Wolf \cite{WOL1959a,RIC1959} constructed a different integral technique based on energy
conservation and the far-field approximation.
The Stratton-Chu and Richards-Wolf methods have been used to study tight
focusing geometries in multiple recent publications
\cite{VAR2000a,LIE2001,QUA2001,BAH2005,BOK2008,APR2008,APR2010b,JEO2015}.
Almost all of these articles have considered simplifications such as
monochromatic light or plane wave incidence, or both, even though they
are not inherent to the Stratton-Chu and Richards-Wolf methods.

Other studies have used a completely different approach and solved Maxwell's
equations with specific source geometries (e.g. complex source beams)
\cite{SHE1999,GON2012} or given four-potential $A_\mu$ \cite{SAL2015} in order to model the
tight focusing regime.
These analytical results are extremely useful in that they provide closed-form approximations
to tightly focused fields. However, they fail to capture the details of the electromagnetic fields
generated by the reflection off a given mirror geometry, as we will show.

In this article, we report on a numerical implementation of the
Stratton-Chu diffraction integrals that fully models the reflection of temporally
short fields off high numerical aperture optical systems. We review the derivation
of the Stratton-Chu equations, noting that the only approximation used to model
the reflection problem is the physical optics approximation.
This approximation does not require any assumption
on the spatial or temporal dependence of the incident field and is valid for almost
any mirror shape. Our specific implementation is thus able to model the reflection of an experimentally
realistic beam profile with a complex broadband spectrum characteristic of the short
pulses produced in high-power laser systems \cite{DAN2015}. In practical terms,
the Stratton-Chu integrals take as an input the laser field impinging on the focusing mirror
and return the focused field at any point in space and time.
This formalism is thus useful to study tightly focused fields, as the unfocused field
coming from the laser chain is typically known to a high experimental accuracy, while the
reflected, tightly focused field is not, due to the difficulty of imaging very
small focal volumes at high intensities.

By construction, the integral representation decouples the regions in which the field
is computed from the region on which we perform the numerical work, which in our case consists of quadratures.
To compute the field in the vicinity of the focal spot of a given mirror, it is therefore
not necessary to model the propagation of the field at every point between
the mirror and the observation point. This confers a distinct numerical advantage
to the integral method compared to the finite-difference time-domain (FDTD),
finite-difference frequency-domain (FDFD) and even traditional finite element methods
(FEM). These latter methods typically discretize the whole region of space
that contains the mirror and the focal spot.  This is especially
taxing in the tight focusing regime where the focal spot must be discretized
with sub-$\lambda$ resolution, while the mirror is several orders of magnitude
larger than $\lambda$. Indeed, laser beams in the optical region of the spectrum
have $\lambda\sim1\,\si{\micro\meter}$ while the mirror has an aperture in the centimeter
range.
This results in a spatial mesh so fine that it would require unmanageable amounts
of memory to accommodate, even on a modern supercomputer. A crude estimate
using a Yee lattice yields a memory requirement approaching the exabyte ($10^6$ terabytes).
This can be mitigated in FEM
techniques by using variable mesh sizes, but this does not remove the inherent
issue that the whole space must be discretized. Moreover, the FDTD, FDFD and FEM
techniques must use an artificial boundary, such as perfectly matched layers \cite{BER1994}, to account for
the free-space propagation after the reflection on the mirror without meshing
the whole space.

Despite its advantages, the versatility of the integral method comes at a cost.
By estimating the number of floating point operations needed to evaluate
the reflected field in a $\lambda^3$ volume around the focal point when a broadband pulse
impinges on a mirror, we can estimate
a runtime of several weeks on a single modern CPU core. Furthermore, even
though the memory requirements are lessened compared to standard methods, they
can still be prohibitive.  Fortunately, the
integrals can be evaluated in parallel very efficiently. This allows for a manageable
runtime, ranging from a few hours to a few days depending on the parameters of
the incident beam, and greatly diminishes the memory required of each CPU core.

The accurate characterization of electromagnetic fields in the strong focusing regime
is essential in the context of the experimental observation of SF-QED effects.
While most studies use simple models, such as plane or paraxial waves, to describe
the laser field while computing SF-QED observables, recent studies have shown that the
spatial structure of the field can alter the observables related to pair
production \cite{FED2009,HEB2011,Schneider2016}, vacuum wave mixing \cite{FIL2015a,Gies2016}
and nonlinear Compton scattering \cite{MAC2011}
in a non-trivial way. Moreover, analytical tools recently developed by Di Piazza
facilitate the study of the effects of tightly focused fields on SF-QED observables \cite{DIP2014,DIP2015,DIP2016}.
Tightly focused fields have also found use in practical applications,
such as direct electron acceleration \cite{SAL2007,VAR2013,MAR2013}, microscopy \cite{YOU2000,LIE2001}
and plasma physics modeling \cite{MAR2013a}.

The article is structured
as follows. Section \ref{sec:theory} contains a
detailed review of the Stratton-Chu representation and their reduction to a
set of diffraction integrals. The physical optics approximation is introduced and shown to arise
from the first term in the Liouville-Neumann expansion.
Section \ref{sec:implementation} discusses practical details of the implementation,
such as the spatial and temporal discretization schemes used, and describes the
efficient parallelization of the algorithm with the domain decomposition method.
It also shows that, for a parabolic
mirror, the integrands of the Stratton-Chu equations do not possess the strong oscillations
typical of diffraction integrals, thus simplifying their evaluation. Section
\ref{sec:analysis} validates the numerical implementation and analyzes the
fields computed via the Stratton-Chu equations and shows that the geometry of
the reflecting surface should be taken into account. Our results show that
future high-power laser facilities could reach the record intensity of $10^{24}\,\si{\watt\per\cm\squared}$,
notwithstanding imperfect vacua and quantum effects.
We conclude in Section \ref{sec:conclu}.


\section{Stratton-Chu diffraction}\label{sec:theory}

This section sets up the theoretical apparatus that will be used to model
the reflection of short optical laser pulses off high numerical aperture (strongly focusing) mirrors.
Starting from Sancer's form of the Stratton-Chu integral representation \cite{SAN1968}, we
derive a set of hypersingular integral equations that describe the fictitious currents
impressed on the mirror by the incident field. We show that the hypersingularity can
be attributed to the openness of the mathematical surface that represents the mirror, as
it reduces to a singular integral in the case of a closed surface. Then, using the fact that the resulting
integral operator is compact, we conclude that the physical optics approximation is
recovered in the first term of the Liouville-Neumann expansion.
Throughout this section, we use Lorentz-Heaviside units in addition to setting
the speed of light to unity, i.e. $c=1$.

\begin{figure}
  \centering
  \includegraphics[width=0.7\columnwidth]{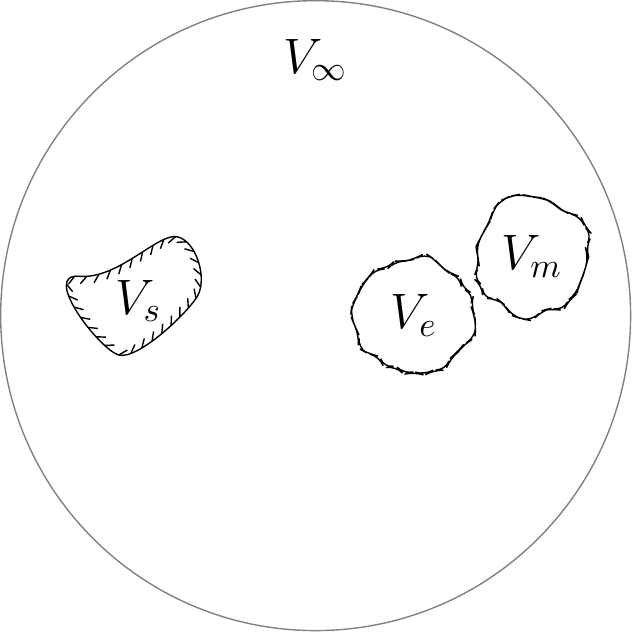}
  \caption{Scattering system used in the derivation of the Stratton-Chu representation.
           $V_e$ and $V_m$ are the support of the electric and magnetic sources, respectively,
           while $V_s$ is the scattering object. The volume $V_\infty$ is a sphere whose radius
           eventually goes to infinity and contains $V_e$ and $V_m$, but not $V_s$.}
  \label{fig:theory.scatteringSystem}
\end{figure}

The Stratton-Chu representation has been derived using a variety of techniques
\cite{KOT1923b,STR1939,POG1973,ROE1992},
but perhaps the most elegant derivation comes from Sancer \cite{SAN1968}.
For the purposes of this work, it suffices to say that the derivation is based
on the application of several vector and tensor calculus identities and the
dyadic-vector version of Green's theorem \cite{TAI1997} on Maxwell's equations.
The geometry of the problem is defined as follows.
Sources of electric and magnetic currents, whose support are $V_e$ and $V_m$, are
supposed to exist in free space, represented by $V_\infty$ (Fig.~\ref{fig:theory.scatteringSystem}).
Mathematically speaking, both $V_e$ and $V_m$ are subsets of $V_\infty$ and can overlap,
and $V_\infty=\mathbb{R}^3\backslash V_s$ is all of free space save from the volume occupied
by the scatterer.
Applying the relevant theorems in this particular geometry yields a representation
in terms of integrals over the surface of the scatterer $S=\partial V_s$.
The representation is valid for surfaces of arbitrary shape, including open surfaces.
The material properties of the scatterer are taken into account
by enforcing proper boundary conditions. The representation, for a monochromatic field,
reads
\cite{SAN1968}
  \begin{subequations}
  \label{eq:theory.field-stratton-chu-sancer}
  \begin{align}
  \bo{E}' &=
     \bo{E}_\text{inc}'+\iint_{S}
    \left\{ ik(\bou{n}\times \bo{B})              g
          +   (\bou{n}\times \bo{E})\times \nabla g\phantom{\frac{i}{k}}\right.\nonumber \\
          & \pushright{\left.+\frac{i}{k}\nabla\nabla g\cdot(\bou{n}\times\bo{B})
    \right\} dS,}
  \label{eq:theory.efield-stratton-chu-sancer}\\
  \bo{B}' &=
    \bo{B}_\text{inc}' + \iint_{S}
    \left\{-ik(\bou{n}\times \bo{E})              g
          +   (\bou{n}\times \bo{B})\times \nabla g\phantom{\frac{i}{k}}\right.\nonumber\\
          &\pushright{\left.-\frac{i}{k}\nabla\nabla g\cdot(\bou{n}\times\bo{E})
    \right\} dS,}
  \label{eq:theory.bfield-stratton-chu-sancer}
  \end{align}
  \end{subequations}
valid $\forall\bo{r}'\in V_\infty$. In these equations, $\{\bo{E}_\text{inc},\bo{B}_\text{inc}\}$
are the incident fields and are assumed
to arise from the sources and currents in $V_e$ and $V_m$, respectively. We thus only
require that the incident fields are solutions of Maxwell's equations. The primed coordinates
represent any point in $V_\infty$, while the unprimed coordinates represent points
on $S$, and are integrated over. $\{\bo{E}', \bo{B}'\}$ thus represent the field at any
given point in $V_\infty$, while $\{\bo{E}, \bo{B}\}$ denote the field on the surface $S$.
The notation $\bo{E}'$ [$\bo{E}$] is short for $\bo{E}(\bo{r}',k)$ [$\bo{E}(\bo{r},k)$]
where $k=2\pi/\lambda$ is the wavenumber of the laser pulse. $g$ is the scalar Green's function
  \begin{equation}
    g(\bo{r},\bo{r}') = \frac{e^{ik|\bo{r}-\bo{r}'|}}{4\pi|\bo{r}-\bo{r}'|}.
  \end{equation}
It is the only factor in Eq.~\eqref{eq:theory.field-stratton-chu-sancer}
that depends on both the integration variables $\bo{r}$
and the observation variables $\bo{r}'$.

Equation (\ref{eq:theory.field-stratton-chu-sancer}) is an integral representation
that expresses the field at any point in $V_\infty$ as integrals of the (yet unknown)
fields $\{\bo{E},\bo{B}\}$ over the surface of the scatterer.
To find the value of the fields on the scatterer, we take the limit $\bo{r}'\rightarrow S$.
This results in an integral equation on $S$, which can be solved iteratively. However,
care must be taken in the evaluation of this limit as $g$ has a simple pole at $\bo{r}=\bo{r}'$,
i.e. on the surface of the mirror.
This singularity is dealt with analytically by deforming the surface at $\bo{r}=\bo{r}'$
to a hemispherical surface with vanishing radius $R$. Let this contour be $S_\epsilon$. The
surface integrals can now be written as
  \begin{align}
    \lim_{\bo{r}'\rightarrow S}\iint_S \left\{\cdot\right\} dS
        &= \lim_{R\rightarrow0}\left(\iint_{S/S_\epsilon}+\iint_{S_\epsilon}\right) \left\{\cdot\right\}dS,
  \end{align}
where $\{\cdot\}$ represents any function. Evaluating this limit for each term separately
yields (the details of the computation are left for Appendix~\ref{sec:app.limits}):
  \begin{subequations}
  \label{eq:theory.singularLimits}
  \begin{align}
    \lim_{\bo{r}'\rightarrow S}\iint_S \bo{A}g dS &= \miint{\rule{1em}{0.5pt}}_S \bo{A}g dS\label{eq:theory.regularLimit}, \\
    \lim_{\bo{r}'\rightarrow S}\iint_S \bo{A}\times\nabla gdS &= \frac{1}{2}\bo{A}\times\bou{n} + \miint{\rule{1em}{0.5pt}} \bo{A}\times\nabla gdS,\label{eq:theory.singularLimit}\\
    \lim_{\bo{r}'\rightarrow S}\iint_S \nabla\nabla g\cdot\bo{A}dS &= \HadamardSurf_S \nabla\nabla g \cdot\bo{A} dS\label{eq:theory.hypersingularLimit},
  \end{align}
  \end{subequations}
where $\bo{A}=\bou{n}\times\bo{F}$ and $\bo{F}$ stands for either the electric or
magnetic field. Equations (\ref{eq:theory.regularLimit}-\ref{eq:theory.singularLimit})
use the Cauchy principal value, denoted by $\CauchySurf$, and Eq.~\eqref{eq:theory.hypersingularLimit}
the Hadamard finite part \cite[Eq. (2.5)]{BLA2000}, denoted by $\HadamardSurf$.
Substituting the results of Eq.~\eqref{eq:theory.singularLimits} in the limit
$\bo{r}'\rightarrow S$ of Eq.~\eqref{eq:theory.field-stratton-chu-sancer}
yields the hypersingular integral equations
  \begin{subequations}
  \label{eq:theory.field-stratton-chu-sancer-hypersingular}
  \begin{align}
  \label{eq:theory.efield-stratton-chu-sancer-hypersingular}
  \bo{E} &= \bo{E}_\text{inc}+ \frac{1}{2}(\bou{n}\times\bo{E})\times\bou{n} \nonumber \\
          &\qquad +\CauchySurf_S
            \left\{ ik(\bou{n}\times \bo{B})              g
                    +   (\bou{n}\times \bo{E})\times \nabla g\right\}dS        \nonumber\\
          &\qquad +\HadamardSurf_S\frac{i}{k}\nabla\nabla g\cdot(\bou{n}\times\bo{B}) dS,\\
  \label{eq:theory.bfield-stratton-chu-sancer-hypersingular}
  \bo{B}
      &= \bo{B}_\text{inc} + \frac{1}{2}(\bou{n}\times\bo{B})\times\bou{n}                  \nonumber\\
      &\qquad+\CauchySurf_S
        \left\{-ik(\bou{n}\times \bo{E})g +(\bou{n}\times \bo{B})\times \nabla g\right\}dS   \nonumber\\
      &\qquad-\HadamardSurf_S\frac{i}{k}\nabla\nabla g\cdot(\bou{n}\times\bo{E})dS,
  \end{align}
  \end{subequations}
for $\bo{r}\in S$. Note that there are no more primed coordinates, as the integral
equations (\ref{eq:theory.field-stratton-chu-sancer-hypersingular})
are valid only for points on the surface of the mirror.

Although it may seem surprising for the
physical electromagnetic fields to be represented by hypersingular integrals,
the singularity is actually caused by the physical discontinuity in the reflecting surface.
The hypersingularity in Eqs.~\eqref{eq:theory.field-stratton-chu-sancer-hypersingular}
disappears if the surface $S$ is closed. Applying Stokes' theorem to the double gradient
term in Eqs.~\eqref{eq:theory.field-stratton-chu-sancer} yields the usual Stratton-Chu
representation for the fields in $V_\infty$
  \begin{subequations}
  \label{eq:theory.field.stratton-chu}
  \begin{multline}
    \bo{E}'   = \bo{E}_\text{inc}'+\frac{1}{ik}\oint_{\partial S} \nabla g \bo{B}\cdot d\bo{\ell}\\
               + \int_S\left\{ ik(\bou{n}\times\bo{B})g
               + (\bou{n}\times\mathbf{E})\times\nabla g
               + (\bou{n}\cdot\bo{E})\nabla g
                 \right\} dS,
 \end{multline}
 \vspace{-0.5cm}
\begin{multline}
    \bo{B}'   = \bo{B}_\text{inc}'-\frac{1}{ik}\oint_{\partial S} \nabla g \bo{E}\cdot d\bo{\ell}\\
              + \int_S\left\{ ik(\bou{n}\times\bo{E})g
               + (\bou{n}\times\mathbf{B})\times\nabla g
               + (\bou{n}\cdot\bo{B})\nabla g
                 \right\} dS.
\end{multline}
\end{subequations}
Taking the limit $\bo{r}'\rightarrow S$ as before reveals that all terms diverge
at most as $1/R^2$. This divergence can be readily integrated using the Cauchy
principal value for the terms inside the surface integral as the integration
measure cancels the divergence. However, this is not true for the terms contained in the line integral,
as the integration measure, $R$, does not cancel the divergence and the integral
is thus hypersingular. It can be shown that, before we take the limit $\bo{r}'\rightarrow S$,
the line integral itself vanishes identically
for a closed surface \cite{SAN1968}, thus removing the hypersingularity.

Let us now go back to Eq.~\eqref{eq:theory.field-stratton-chu-sancer-hypersingular} and
impose the appropriate boundary conditions for a perfectly
conducting mirror \cite[Eq. (1.18)]{STR1941}, i.e.
  \begin{equation}
    \label{eq:theory.boundaryConditions}
    \bou{n}\times\bo{E}=0;\qquad \bou{n}\times\bo{B}=\bo{J}.
  \end{equation}
Extracting the tangential components of Eqs.~\eqref{eq:theory.field-stratton-chu-sancer-hypersingular}
and imposing these conditions yields
  \begin{subequations}
  \begin{align}
    -\bou{n}\times\bo{E}_\text{inc} &= \bou{n}\times\left[\CauchySurf_S ik\bo{J}g dS + \HadamardSurf_S\frac{i}{k}\nabla\nabla g\cdot\bo{J} dS\right],\\
    \label{eq:theory.mfie}
    \frac{1}{2}\bo{J}  &= \bo{J}_\text{inc} + \bou{n}\times\CauchySurf_S \bo{J}\times\nabla gdS,
  \end{align}
  \end{subequations}
where $\bo{J}_\text{inc}=\bou{n}\times\bo{B}_\text{inc}$.
Both equations can be solved for the current $\bo{J}$ induced by the incident
field $\{\bo{E}_\text{inc},\bo{B}_\text{inc}\}$. Since it can be shown that
the integral operator in the magnetic field integral equation, \eqref{eq:theory.mfie},
is compact, we can use the Liouville-Neumann series to solve the integral equation
iteratively \cite[\S6.16]{JON1994}. The first term of the series yields the usual
physical optics approximation (POA)
  \begin{equation}
    \label{eq:theory.poa}
    \bo{J} = 2\bo{J}_\text{inc},
  \end{equation}
which coincides with the result for a plane, infinite mirror \cite[\S12.2]{BLA2007}. The
integral in Eq.~\eqref{eq:theory.mfie} can thus be interpreted as a curvature effect.
Indeed, each term in the iterative solution can be shown to get gradually smaller
in magnitude if the radius of curvature is larger than the wavelength of the
incident radiation \cite{CUL1958,BLA2007}.
The POA has been used successfully in many studies \cite{BOU1954,LOV1978}.

Substituting the POA [Eq.~\eqref{eq:theory.poa}] and the boundary conditions
[Eq.~\eqref{eq:theory.boundaryConditions}] in Eqs.~\eqref{eq:theory.field-stratton-chu-sancer},
we can express the reflected field $\bo{F}'_\text{ref}=\bo{F}'-\bo{F}_\text{inc}$
in $V_\infty$ as
  \begin{subequations}
  \label{eq:theory.stratton-chu-poa}
  \begin{align}
    \label{eq:theory.stratton-chu-poa.efield}
    \bo{E}'_\text{ref}  &=  2\iint_S
      \left\{
        ik(\bou{n}\times\bo{B}_\text{inc})g
        +\frac{i}{k}\nabla\nabla g\cdot(\bou{n}\times\bo{B}_\text{inc})
        \right\}dS, \\
    \label{eq:theory.stratton-chu-poa.bfield}
    \bo{B}'_\text{ref}  &= 2\iint_S (\bou{n}\times\bo{B}_\text{inc})\times\nabla g dS.
  \end{align}
  \end{subequations}
Even though Eqs.~\eqref{eq:theory.stratton-chu-poa} are valid expressions
for the reflected field, the double gradient term tends to strongly oscillate in applications and therefore
make its numerical evaluation difficult. We sidestep this issue by once again applying
Stokes' theorem on the double gradient term, which leads to
  \begin{subequations}
  \label{eq:theory.stratton-chu-poa-og}
  \begin{align}
    \label{eq:theory.stratton-chu-poa-og.efield}
    \bo{E}'_\text{ref}(\bo{r}',k)  &= 2\iint_S
      \left\{
        ik(\bou{n}\times\bo{B}_\text{inc})g
        +\left(\bou{n}\cdot\bo{E}_\text{inc}\right)\nabla g
        \right\}dS \nonumber\\
            &- \frac{2}{ik}\oint_{\partial S}\nabla g
                    \left[\bou{n}\times(\bou{n}\times\bo{B}_\text{inc})\right]\cdot d\bo{\ell},\\
    \label{eq:theory.stratton-chu-poa-og.bfield}
    \bo{B}'_\text{ref}(\bo{r}',k)  &= 2\iint_S (\bou{n}\times\bo{B}_\text{inc})\times\nabla g dS.
  \end{align}
  \end{subequations}
The double gradient term has been replaced by two terms: a surface term with a
single gradient and an additional line integral term  which also contains
a single gradient. The single gradient results in a $1/R$ behavior of the integrands,
compared to the $1/R^2$ of the double gradient. This weakens the oscillations of
the integrands.

Equations (\ref{eq:theory.stratton-chu-poa-og}) describe a spectral component
of frequency $k$ of the reflected field at a given position $\bo{r}'$ as an integral
of the incident field on the surface of the mirror. The remainder of this paper
will be devoted to their efficient numerical evaluation for arbitrary incident
fields $\{\bo{E}_\text{inc},\bo{B}_\text{inc}\}$ with complex time-dependence
and for arbitrary mirror geometries $S$.

\section{Practical Implementation}\label{sec:implementation}

This section discusses the numerical evaluation of the integrals in
Eqs.~\eqref{eq:theory.stratton-chu-poa-og} in some detail. First, we discuss the projection
of the incident field on the mirror and introduce the cylindrical coordinate system used
throughout the article. Second, we discuss the scale separation that the integral
method exhibits, and the domain decomposition strategy that it allows. We also discuss
the spatial and temporal discretization schemes used in the numerical implementation.
Third, we show that for a specific mirror geometry, the paraboloid mirror, the integrands
do not exhibit the strong oscillations typical of diffraction integrals, allowing
the use of simple quadrature methods.

\begin{figure}
  \centering
  \includegraphics{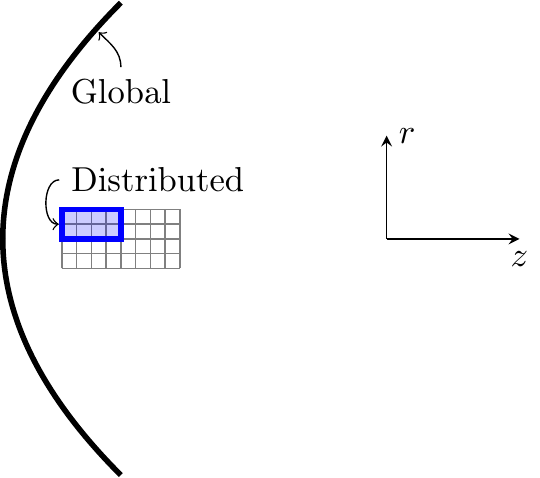}
  \caption{Domain decomposition method applied to the mesh in the focal region
           (not to scale). The mesh over the mirror is global, i.e. all processors
           have a copy of the whole mesh. The mesh in the focal region
           is distributed over a number of processors. Each processor carries
           a portion of the whole mesh, and computes the field values for this
           portion of the mesh only. The blue region depicts the portion that a single
           processor computes. A cylindrical coordinate system is used.}
  \label{fig:impl.domain-decomposition}
\end{figure}

\subsection{Incident Field Models}
The first step in computing the reflected field in the focal region is to specify the
incident field $\{\bo{E}_\text{inc},\bo{B}_\text{inc}\}$, i.e.~the field that
impinges on the mirror $S$. Since the Stratton-Chu formalism does not restrict
it in any way, it would in principle be possible to measure the amplitude and
phase of each frequency component of the electromagnetic
field on the surface of the mirror and use it as an input parameter of
Eqs.~\eqref{eq:theory.stratton-chu-poa-og}. However, obtaining a complete cartography
of the incident field is experimentally very challenging, and it will be more useful
to opt for a different approach.

To determine the value of the field on the mirror, we use the Gauss-Laguerre modes
of the paraxial wave equation as an expansion basis to describe both linearly and
radially polarized beams (see Supplementary Material for explicit expressions).
The use of the paraxial approximation is justified in the description of the
incident beam as it incurs an error on the order of the divergence angle, i.e. $\mathcal{O}(\epsilon)$.
For a highly collimated laser beam, we have $\epsilon\ll1$. Moreover, since the
Stratton-Chu equations are linear, the error in the reflected field will also be
$\mathcal{O}(\epsilon)$, even though the reflected field could be tightly focused.
The use of these closed-form expressions allows us
to easily and quickly evaluate the integrands of Eqs.~\eqref{eq:theory.stratton-chu-poa-og}
regardless of the discretization procedure used on the mirror. However, this approach
neglects the diffraction of the incident field by the aperture of the mirror. This
is justified as the aperture size, $r_\text{max}$, is orders of magnitude larger than the wavelength
of the incident beam, i.e. $r_\text{max}\gg\lambda$ in typical optical experiments.

Since we will consider mostly axisymmetric mirror sha\-pes, such as paraboloid
mirrors, we use a cylindrical coordinate system (Fig.~\ref{fig:impl.domain-decomposition}).
The incident field is assumed to be propagating in the $-z$ direction towards
the mirror. For simplicity, and without loss of generality, we assume that the
mirror is described by the explicit parametrization $z=F(r,\theta)$.
Explicit expressions of Eqs.~\eqref{eq:theory.stratton-chu-poa-og} in these conditions
are shown in the Supplementary Material.

\subsection{Temporal Discretization}
\label{sec:impl.temporalDist}

In deriving Eqs.~\eqref{eq:theory.stratton-chu-poa-og}, we have implicitly assumed that
the incident field had been Fourier-transformed, i.e.
  \begin{equation}
    \bo{F}(\bo{r},t) = \int_{-\infty}^\infty \bo{F}(\bo{r},k)e^{-ikt}dk
  \end{equation}
where $k=\omega$ is the wavenumber of the radiation.
In the numerical implementation
the discrete version of the transform is used, i.e.
  \begin{equation}
    \label{eq:impl.semiDiscreteFFT}
    \bo{F}(\bo{r},t) = \sum_{n=-N_\text{max}}^{N_\text{max}} \bo{F}_n(\bo{r},k_n)e^{-ik_n t}\Delta k
  \end{equation}
where $\bo{F}_n$ is the $n$th frequency sample, or component, $k_n$ the sampled
frequency and $\Delta k$ the interval between samples. This discrete version is
periodic. $N_\text{max}$ and $\Delta k$ must be chosen such that the entire duration of the pulse
can be represented with Eq.~\eqref{eq:impl.semiDiscreteFFT} and the Nyquist criterion
must be obeyed.
To account for the time dependence
of the reflected field, the evaluation of the integrals in Eqs.~\eqref{eq:theory.stratton-chu-poa-og}
must be performed for each frequency component $\bo{F}_n$.

To determine the relative amplitude of each component, we normalize them using a given
power spectrum of fixed energy. This is sufficient to uniquely determine the
time dependence of the incident beam.
Experimentally, it is possible to measure the energy contained in each component $n$.
Using the Poynting theorem, we can show that the energy density in each mode is
given by the integral of the Poynting vector over an infinite plane, i.e.
  \begin{equation}
    \epsilon(k_n) = 4\pi\oiint_A\real{\bo{E}_n(\bo{r},k_n)\times\bo{B}_n^*(\bo{r},k_n)}\cdot d\bo{A}.
  \end{equation}
Since Maxwell's equations are linear, each frequency component $\bo{E}_n$ can be
renormalized as
  \begin{equation}
    \bo{E}_n = E_0^n \bo{f}_E^n; \qquad \bo{B}_n = E_0^n \bo{f}_B^n,
  \end{equation}
where $E_0^n$ the arbitrary amplitude of the component and $\bo{f}_E\,(\bo{f}_B)$
contain the spatial dependence
of the field \footnote{While these functions are arbitrary, we evaluate this integral
for specific beam models in the Supplementary Material.}.
This allows us to write
  \begin{equation}
    \label{eq:theory.normalizationAmplitude}
    E_0^n = \sqrt{\frac{\epsilon(k_n)}
                {4\pi\oiint_A\real{\bo{f}_E(\bo{r},k_n)\times\bo{f}_B^*(\bo{r},k_n)}\cdot d\bo{A}}}
  \end{equation}
where $\epsilon(k_n)$ is a known power spectrum, normalized such that
  \begin{equation}
    \label{eq:theory.totalEnergy}
    E_\text{tot} = \int_{-\infty}^\infty \epsilon(k)dk
  \end{equation}
where $E_\text{tot}$ is the total energy of the beam incident on the mirror.

An arbitrary spectral phase relation can be enforced simply by mapping
  \begin{equation}
  \begin{aligned}
    \bo{E}_\text{inc}(\bo{r},\omega)&\mapsto \bo{E}_\text{inc}(\bo{r},\omega)e^{i\phi(\omega)},\\
    \bo{B}_\text{inc}(\bo{r},\omega)&\mapsto \bo{B}_\text{inc}(\bo{r},\omega)e^{i\phi(\omega)},
  \end{aligned}
  \end{equation}
where $\phi(\omega)$ is any function of the frequency. As an example, chirped
mirrors can give rise to a frequency dependent phase \cite{JEO2015}.

To ensure an efficient reconstruction of the time-depen\-dent field, the reflected
field should be kept in RAM for the entirety of the computational process. While
this implies a larger memory use, it ensures that we do not need to read data from
to the disk in the evaluation of the semi-discrete Fourier transform, Eq.~\eqref{eq:impl.semiDiscreteFFT}.

Note that this approach of modeling the time dependence of the field neglects
any spatio-temporal couplings \cite{PAR2016}. However, the implementation could
be updated to support more general time-dependence, such as the position-dependent
power spectra that characterizes spatio-temporal couplings.

\subsection{Spatial Discretization}
The evaluation of Eqs.~\eqref{eq:theory.stratton-chu-poa-og} demands the definition of
two separate spatial domains (Fig.~\ref{fig:impl.domain-decomposition}).
The first is the mirror $S$ on which the incident field
is assumed known. The second is the region of space in which we wish to compute
the reflected field. Since we are mostly interested in the behavior of the
reflected field in the vicinity of the focal spot, we will denote this domain
the focal region. In the numerical implementation, these two domains are discretized
separately. This naturally separates the two length scales that appear in the evaluation
of Eqs.~\eqref{eq:theory.stratton-chu-poa-og}, as we show in Sec.~\ref{sec:impl.domainDecomp}.

\subsubsection{Meshing the Mirror: Integrand Analysis}
\label{sec:impl.integrands}

Given the form of Eqs.~\eqref{eq:theory.stratton-chu-poa-og},
  \begin{equation}
    \label{eq:impl.oscillatingIntegrand}
    I = \int f(\bo{r})e^{iku(\bo{r},\bo{r}')}d\bo{r}
  \end{equation}
where $u(\bo{r},\bo{r}')=|\bo{r}-\bo{r}'|$ is the oscillatory part of the Green's
function, one can conclude that the discretization needs to be fine enough
to resolve the oscillations of the complex exponential. At first glance, this integrand seems to
oscillate rapidly because of the large exponent $k|\bo{r}|\gg1$. Indeed,
$r$ scales with the size of the parabola $\mathcal{O}(\si{\cm})$, while
$k\sim\mathcal{O}(\si[per-mode=reciprocal]{\per\micro\metre})$ for optical pulses. Rapidly oscillating
integrands of the form of Eq.~\eqref{eq:impl.oscillatingIntegrand} are difficult
to evaluate numerically, although there are some techniques
that facilitate their evaluation \cite{ISE2004,GAN2007}. Here, however, we show
that for a parabolic mirror, a non-trivial cancellation in the phase function
results in the strength of the oscillations being controlled by the observation variables ($k|\bo{r}'|\sim1$)
rather than by the integration variables ($k|\bo{r}|\gg1$). It is thus not necessary
for the discretization scheme on the mirror to resolve the wavelength, only the variations
in the shape of the incident fields $\bo{f}_E$ and $\bo{f}_B$.
This allows the use of simple quadrature methods, such as tensor products of
either Simpson \cite[Eq. (4.1.14)]{PRE1989} or Gauss-Legendre \cite[\S4.6]{PRE2007}.

The cancellation occurs when we consider the oscillatory behavior of the incoming
field. The paraxial fields that are used to represent the incident laser pulses generically have
$f(\bo{r})\sim e^{-ikz}$ behavior (see Supplementary Material).
Hence, the function that needs to be integrated
in the exponential above is not $u(\bo{r}-\bo{r}')$, but rather $u(\bo{r}-\bo{r}')-z$.
If this function can be shown to be small, i.e. $ku(\bo{r},\bo{r}')-kz\ll1$, then
the integrands do not have strong oscillations. We examine the phase function
in the vicinity of the focal spot, i.e. around $\bo{r}'=0$. We can
write the phase function as a bidimensional Taylor series
  \begin{multline}
    u(\bo{r},\bo{r}')-z = \left(\sqrt{r^2+z^2}-z\right) \\- \frac{r\cos(\theta-\theta')}{\sqrt{r^2+z^2}} r'
                                              - \frac{z}{\sqrt{r^2+z^2}}z'
                        + \mathcal{O}\left(\frac{r',z'}{|\bo{r}|}\right).\label{eq:impl.taylorPhase}
  \end{multline}
Near the focal spot, then, the first-order terms oscillate slowly, because the prefactor
of each term has range $[-1,1]$ and $r'$ and $z'$
(the radial and longitudinal distances from the focal point)
do not exceed a few wavelengths, i.e. $k|\bo{r}'|\sim1$.
The zeroth-order term (parenthesis in Eq.~\eqref{eq:impl.taylorPhase})
is thus the only term that can lead to a rapidly oscillating
integrand.
For the specific of a parabolic mirror, for which $z=r^2/4f-f$, the phase function
reads
  \begin{equation}
    u(\bo{r},\bo{r}')-z \simeq 2f - \frac{4rf}{f^2+r^2} r' - \frac{r^2-4f^2}{r^2+4f^2} z' + \cdots
  \end{equation}
The zeroth-order term does not depend on the integration variables, implying
that the integrand does not oscillate at $\bo{r}'=0$ and oscillates weakly
in its vicinity.

\begin{figure}
  \centering
  \includegraphics{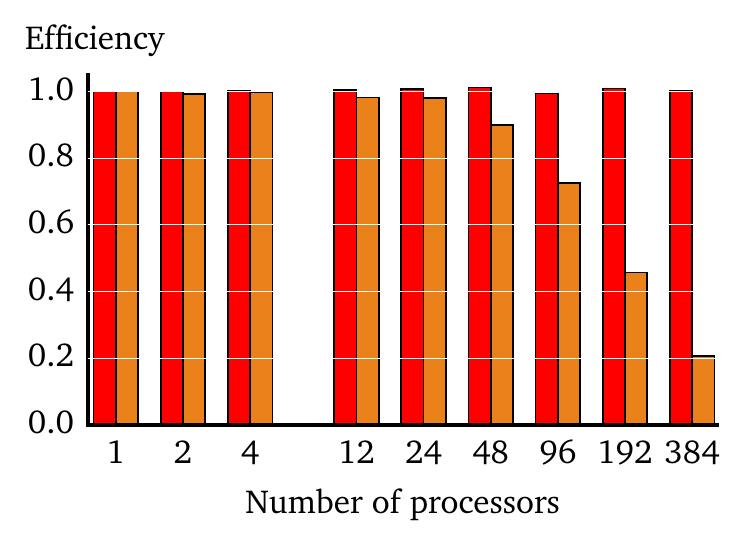}
  \caption{Parallel efficiency of the algorithm without data output (red) and with data
           output (orange). If run without data output,
           the implementation remains very efficient regardless of the number of
           processors (384 is the maximum that can be requested on the supercomputer on
           which this test was run). However, the efficiency dramatically drops above
           24 processors if there is data output (orange bars). This is due to slow communication
           between nodes and is not inherent to the algorithm.}
  \label{fig:impl.parallel-efficiency}
\end{figure}

This non-trivial cancellation does not carry over to generic surfaces, however.
For instance, the zeroth-order term does depend on the integration variables
in the ellipsoidal \footnote{More precisely,
mirrors of the form $z^2/c^2=1-(x^2+y^2)/a^2$, with focal spots
at $z=\pm\sqrt{c^2-a^2}$.} case. In fact, for other mirror geometries, the initial
physical motivation of looking at distances not too far away from the focal spot
might be lacking. For most mirrors, there are no uniquely
defined focal points and the field is diffuse. In these cases, care
should be taken when numerically evaluating the Stratton-Chu integrals.

We will use a parabola as the mirror geometry for the remainder of this paper,
as it provides the highest field intensities and simplifies the quadrature procedure.

\subsubsection{Domain Decomposition}
\label{sec:impl.domainDecomp}
While the mirror does not need to be discretized with sub-$\lambda$ resolution,
the focal region requires it.
In the tight focusing regime,
the focal spot is approximately the size of the incident wavelength, namely
$\lambda\sim1\,\si{\micro\metre}$ for optical lasers, while the mirror and
the spatial extension of the incident field usually have centimeter length scales.
As hinted on in the introduction, this results in unmanageable memory requirements
in traditional computational methods, as they typically need to mesh the entire
region of space between the parabola and the focal spot in order to propagate
the field between these two regions.

In the Stratton-Chu
formalism, only the focal region and the mirror need to be discretized, considerably lowering
the memory requirements. However, they can still be prohibitive, requiring
a few gigabytes (GBs) per frequency component. This issue is compounded by the fact that
temporally short pulses have broad spectra, requiring us to compute
Eqs.~\eqref{eq:theory.stratton-chu-poa-og}
for a large number of frequency components. Since they must be kept in memory
to reconstruct the temporal field, this can lead to data outputs that are approximately
100GB in size, for a typical mesh size of $150\times150\times150$ points in the
focal volume, with $r\in[0,2.5]\,\si{\micro\metre}$, $\theta\in[0,2\pi]$ and
$z\in[-2.5,2.5]\,\si{\micro\metre}$. This number may vary significantly as per the incident field
model, the number of frequency samples and the strength of the focusing optics.

To mitigate this issue, it is necessary to evaluate Eqs.~\eqref{eq:theory.stratton-chu-poa-og}
in parallel.
Since the fields at different points in
space are computed independently of each other, this can be done efficiently.
To achieve maximum efficiency, we use the domain decomposition strategy.
Once the focal region has been meshed to the desired accuracy, the
mesh is divided onto multiple processors. Each processor contains a portion
of the total mesh in its assigned memory, but contains the entirety of the
mesh of the mirror (Fig.~\ref{fig:impl.domain-decomposition}). This way, each processor
evaluates the integrals of Eqs.~\eqref{eq:theory.stratton-chu-poa-og}
for its portion of the focal spot mesh independently of the others and for each
frequency component sequentially. Once the
computation is complete, the fields can be written to disk in parallel.


\subsubsection{Parallel Efficiency and Runtimes}

We have implemented this algorithm in C++ using the open source OpenMPI parallel
library. The output is also parallelized using the parallel version of the HDF5
library. The implementation is efficient (Fig.~\ref{fig:impl.parallel-efficiency}).
The efficiency is defined by the metric
  \begin{equation}
    E_n = \frac{t_1}{nt_n}
  \end{equation}
where $n$ is the number of processors and $t_n$ is the time it takes to run
the program with $n$ processors. Without data output, the algorithm has essentially
a perfect parallel efficiency even at a high number of processors ($N_\text{max}=384$). We surmise,
given the embarrassingly parallel nature of the algorithm, that the implementation
could scale to a higher number of processors. 384 is the maximum number
that can be requested on the supercomputer that is available to our group.
The data output, however, is a point of contention. The parallel efficiency drops dramatically
when using more than 24 processors (Fig~\ref{fig:impl.parallel-efficiency}).
This is caused by the fact that each node
(a group of processors that share the same memory) possesses 24 processors.
When run on more than 24 processors, the data output function is required to
communicate between different nodes that are linked via InfiniBand cables. This
bottleneck dramatically increases the runtime of the HDF5 output facilities
and thus reduces the parallel efficiency.
Note that this is simply a limitation of the cluster's memory architecture,
not of our implementation or of the HDF5 library.

In this section, we have discussed the memory requirements and execution
times of our algorithm when evaluating the electromagnetic field in a volume
around the focal point and for a large number of temporal slices. It is worth putting
these numbers into perspective by discussing the cost of evaluating the field
at a single point in space, and for a single frequency component, i.e. the cost
of a single numerical quadrature over the focusing mirror. For a typical
mesh size of $250\times250$ points on the reflecting surface, this evaluation takes approximately
$0.3\,\si{\second}$ on a single core of an Intel Core i7-4700MQ CPU.
The computation time scales linearly with
the number of points used on the mesh of the mirror and also with
the number of frequency components. If one needs to compute
the field at a given point in time, all field components should be computed at
the same spatial point and the semi-discrete Fourier transform should be taken. Usually,
the Fourier transform takes a negligible amount of time compared to the field calculation.
For a typical optical $20\,\si{\femto\second}$ pulse, 50 frequency components suffice
to accurately sample the power spectrum and computing the field at a single
space-time point takes $50\times0.3\,\si{\second}=15\,\si{\second}$.

\section{Fields in the focal region}\label{sec:analysis}
The unavailability of closed-form solutions of Eq.~\eqref{eq:theory.stratton-chu-poa-og}
makes a systematic verification of any numerical implementation of the
Stratton-Chu diffraction integrals difficult. However, since they represent physical fields,
the values calculated with our implementation should obey Maxwell's equations
and should satisfy the principle of energy conservation.
We verify that the reflected fields computed via our implementation converge
as we reduce the discretization size, and that they obey the previously listed conditions.

\subsection{Numerical Checks}
In this section, we study the convergence of the evaluation of the integrals
in Eqs.~\eqref{eq:theory.stratton-chu-poa-og}. We also verify that the computed fields
obey Maxwell's equations in both the frequency and time domains and that our
implementation conserves the energy of the system.

We use a radially polarized Gaussian beam as the incident
field (see Supplementary Material for the explicit expressions).
We also assume that its spectral power is super-Gaussian in $\lambda$ and that
the full-width half-maximum (FWHM) temporal duration associated with the spectrum
is $18\,\si{\fs}$ (see Table \ref{tab:comparison.parameters} for detailed parameters).
The parabolic mirror has a numerical aperture of $\text{NA} = 1$,
i.e. $r_\text{max}=2f$. The transverse width of the incident field is chosen such that
the field is not clipped by the edge of the mirror.
While we present our test results only for the radial polarization
to save on space, we have verified that the results hold also for linearly polarized
fields.

\begin{figure}
  \centering
  \includegraphics[width=\columnwidth]{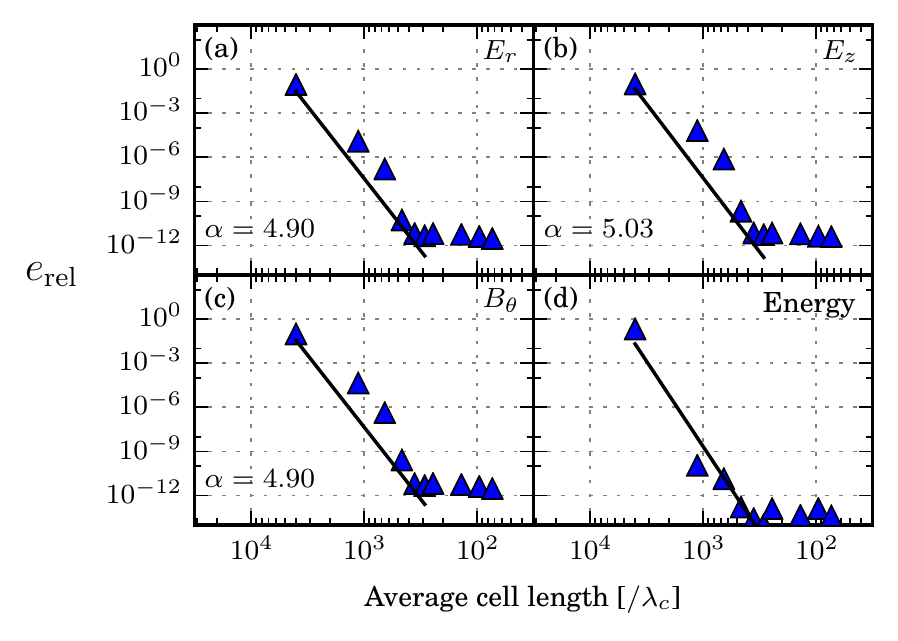}
  \caption{(a)--(c) Relative error $e_\text{rel}$ of each component of the radially polarized
           electromagnetic field in the focal spot (blue triangles) as a function of the average polar
           cell length given as a fraction of the central wavelength $\lambda_c$. The slope $\alpha$ of
           the linear fit (black curves) shows that the convergence order is given by the order
           of the quadrature method for each component. (d) Relative difference between
           the energy of the incident beam and the energy of the field in the focal spot.
           The linear fit (black curve) shows a convergence order consistent with
           the energy being a quadratic function of the fields. See Table \ref{tab:comparison.parameters}
           for the simulation parameters used to generate this figure.}
  \label{fig:comparison.convergenceComponents}
\end{figure}

To verify that our results have converged, we evaluate the reflected field, i.e. the surface integrals
in Eqs.~\eqref{eq:theory.stratton-chu-poa-og}, for different mesh sizes. We then use
the results obtained with the finest mesh as a reference to measure the convergence
speed. To do so, we compute the relative error, defined as the maximum value of
the absolute difference of the fields over the focal region divided by the
maximum value of the magnitude of the reference field
  \begin{equation}
    e_\text{rel} = \frac{\max_{\bo{r}'} \left|\bo{F}_{N_\text{max}}-\bo{F}_N\right|}
                        {\max_{\bo{r}'} \left|\bo{F}_{N_\text{max}}\right|}
  \end{equation}
where $\bo{F}_N$ is the electromagnetic field computed with $N$ radial discretization points.
Because we use cylindrical coordinates, the cells of the mesh increase in area with
the distance from the origin. We use the average area,
  \begin{equation}
    \left\langle A \right\rangle_N = \frac{\Delta\theta(\Delta r)^2}{2}\sum_{n=0}^{N}(2n+1)
                                   = \frac{\Delta\theta(\Delta r)^2}{2}\frac{(N+1)^2}{N},
  \end{equation}
where $\Delta\theta$ is the angular mesh resolution,  $\Delta r$ the radial mesh resolution
and $N$ of radial mesh points, as a measure of the mesh resolution.
Each individual component converges according to the order
of the quadrature method that was used to evaluate the integrands.
Specifically, we
used a fifth order method, and the components converge following $h^{-\alpha}$
with $\alpha\sim5$ (Fig.~\ref{fig:comparison.convergenceComponents}a-c).

It is interesting to note
that the typical mesh size necessary for convergence (the saturated portions of
Fig.~\ref{fig:comparison.convergenceComponents}) is more than a hundred (100) times larger than the wavelength of the
beam. This is a direct consequence of the results of Sec.~\ref{sec:impl.integrands}.

Our implementation also satisfies the principle of conservation of energy and
its numerical value converges following the order of the quadrature method,
same as the individual components
(Fig.~\ref{fig:comparison.convergenceComponents}d). To verify this, we have
computed the total energy in the domain given by (in SI units)
  \begin{equation}
    E = \frac{1}{2}\iiint_V \left[\epsilon_0\bo{E}(\bo{r},t)^2+\frac{1}{\mu_0}\bo{B}(\bo{r},t)^2\right]dV,
  \end{equation}
as a function of the average mesh size.
This quantity is equal to the integral of the energy density in the frequency domain,
Eq.~\eqref{eq:theory.totalEnergy}.

To ensure that the electromagnetic energy contained in the focal
region and the energy of the incident beam are numerically equal,
we must make sure that the spatial extension of the incident
beam is smaller than the aperture of the mirror so it is not clipped.
It is also necessary to compute the reflected field in a region of space that spans
several wavelengths, on the order of $25\lambda$ in each direction. This is substantially larger
than the $\lambda^3$ volume we are usually interested in. This is because
the longitudinal components have a larger spatial extension than the transverse
ones (see next section for a discussion).
In turn, the larger focal region forces us to use a finer mesh on the parabola, as the
integrands oscillate more strongly at larger distances from the geometrical focal
point, as discussed in Sec.~\ref{sec:implementation}.

\begin{figure}
  \centering
  \includegraphics[width=\columnwidth]{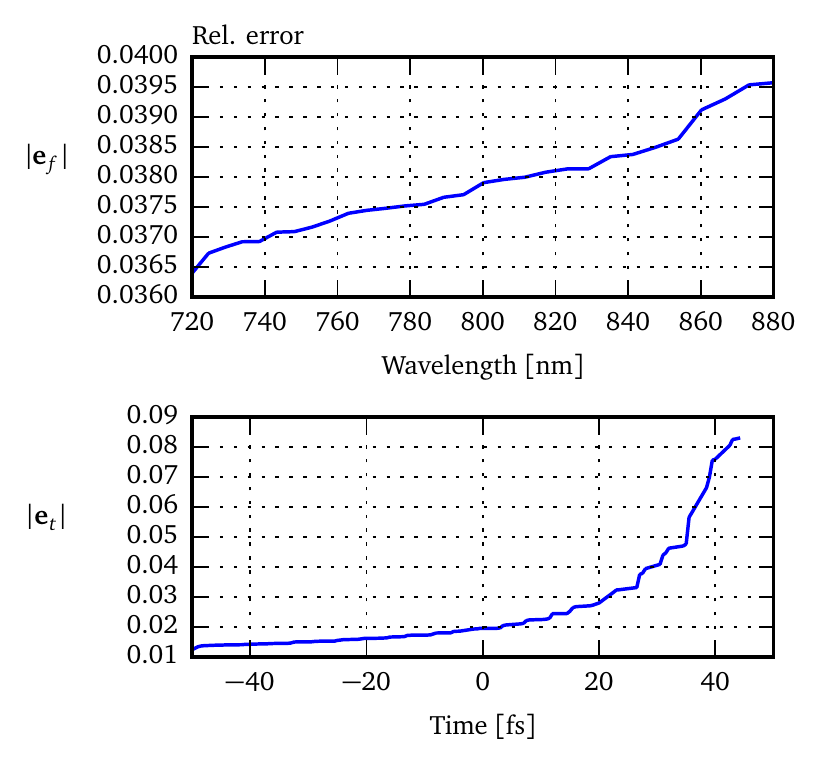}
  \caption{Average value of the magnitude of the relative error vectors $\bo{e}_f$ and $\bo{e}_t$ over
           the focal region.}
  \label{fig:comparison.maxwellEquations}
\end{figure}

The fields in the focal spot have numerically been shown to obey Maxwell's
equations, in both the frequency and time domains (Fig.~\ref{fig:comparison.maxwellEquations}).
To show this, we computed the magnitude of the relative difference vectors
  \begin{subequations}
  \label{eq:comparison.relDiffMaxwell}
  \begin{multline}
    \label{eq:comparison.relDiffMaxwellFreq}
    \bo{e}_\text{f} =
      \frac{2(\nabla\times\bo{E}-ik\bo{B})}{\max_{\bo{r}'}{|\nabla\times\bo{E}|}+\max_{\bo{r}'}{|ik\bo{B}|}}\\
      +\frac{2(\nabla\times\bo{B}+ik\bo{E})}{\max_{\bo{r}'}{|\nabla\times\bo{B}|}+\max_{\bo{r}'}{|ik\bo{E}|}},
  \end{multline}
  \begin{multline}
    \label{eq:comparison.relDiffMaxwellTime}
    \bo{e}_\text{t} =
      \frac{2(\nabla\times\bo{E}+\partial_t\bo{B})}{\max_{\bo{r}'}{|\nabla\times\bo{E}|}+\max_{\bo{r}'}{|\partial_t\bo{B}|}}\\
      +\frac{2(\nabla\times\bo{B}-\partial_t\bo{E})}{\max_{\bo{r}'}{|\nabla\times\bo{B}|}+\max_{\bo{r}'}{|\partial_t\bo{E}|}},
  \end{multline}
  \end{subequations}
which should vanish if the fields obey Maxwell's equations. The curls and time derivative are computed
via a central finite difference scheme on the mesh. This specific definition for the
relative difference is chosen as to provide a natural scale for the quantities
in the numerator of Eq.~\eqref{eq:comparison.relDiffMaxwell}.

Equation (\ref{eq:comparison.relDiffMaxwellFreq}) tests the Stratton-Chu quadrature
routines that evaluate Eqs.~\eqref{eq:theory.stratton-chu-poa-og}, as we perform
all in our calculations in the frequency domain.
Equation (\ref{eq:comparison.relDiffMaxwellTime}) tests the temporal reconstruction
routine, i.e. our implementation of Eq.~\eqref{eq:impl.semiDiscreteFFT}.

\subsection{Tightly Focused Linearly Polarized Gaussian Beam}

\begin{figure*}
  \includegraphics[width=\textwidth]{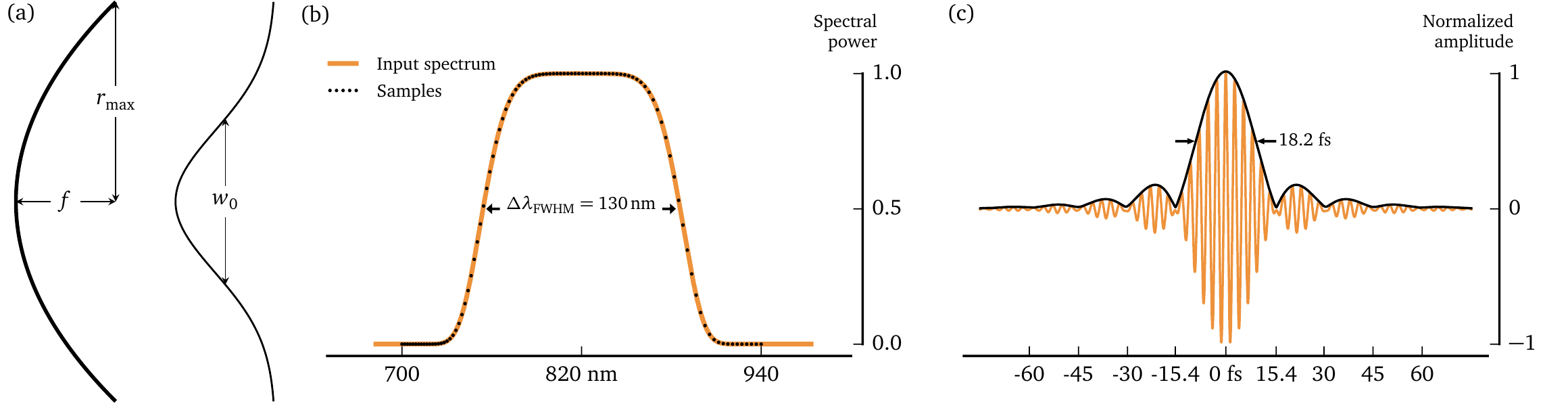}
  \caption{Characteristics of the incident beam. \textbf{(a)} A Gaussian beam with a beam waist
           $w_0$ impinges on parabolic mirror of focal length $f$ and aperture size $r_\text{max}.$
           \textbf{(b)} The beam has a super-Gaussian power spectrum centered at
           $\lambda_c=820\,\si{\nano\metre}$ and has a full width at half maximum (FWHM) of
           $130\,\si{\nano\metre}$. We used 100 frequency samples (black dots) in the numerical simulation.
           \textbf{(c)} In the time domain, this corresponds to a field that has a main pulse with
           FWHM duration of $18.2\,\si{\fs}$ and weaker, shorter revivals at earlier and later times.}
  \label{fig:comparison.parabolicMirror}
\end{figure*}

In this section, we show a concrete example of a calculation that can be
performed using our efficient implementation of the Stratton-Chu
integral representation. We study the spatio-temporal focusing
of a linearly polarized (along the $x$-axis), paraxial femtosecond pulse incident onto a high numerical aperture
on-axis parabola.  A similar setup was first used in a scheme to directly accelerate electrons
with a high-intensity radially polarized field \cite{PAY2012,MAC2012,PAY2013_pat,LAC2015}.

The relevant parameters of the simulation are defined in Fig.~\ref{fig:comparison.parabolicMirror}.
The shape of the parabola is fixed by its focal length $f$ and its aperture size
$r_\text{max}$. The numerical aperture is determined geometrically from the opening
angle of the parabola and is given by $\text{NA}=\sin\theta$. In
Fig.~\ref{fig:comparison.parabolicMirror}(a), the parabola has $\text{NA}=1$.
The power spectrum, shown in Fig.~\ref{fig:comparison.parabolicMirror}(b), is chosen to have a
super-Gaussian shape
  \begin{equation}
    \epsilon(\lambda) \propto \exp\left[\left(\frac{\lambda-\lambda_c}{\Delta\lambda}\right)^{2n}\right],
  \end{equation}
which corresponds to the time dependence pictured in Fig.~\ref{fig:comparison.parabolicMirror}(c).
The numerical values of the parameters were chosen as to best approximate the specifications of
planned high-power laser facilities \cite{PAP2013,LEB2013,SPE2014} and are shown in
Table \ref{tab:comparison.parameters}.

\begin{table}
  \centering
  \begin{tabular*}{\columnwidth}{@{\extracolsep{\fill} }lS[table-number-alignment = center, table-align-exponent = false]c|lSc}
    \toprule
    \multicolumn{6}{c}{\textbf{Simulation Parameters}} \\
    \midrule
    \multicolumn{3}{c|}{Parabola}   & \multicolumn{3}{c}{Incident Beam}\\
    Param.        & {Value}     & Unit        & Param.       & {Value}     & Unit                 \\
    \midrule
    $r_\text{max}$   & 0.125       & \si{\metre} & $w_0$           & 0.100       & \si{\metre}          \\
    $f$              & 0.0675      & \si{\metre} & $\lambda_c$     & 820         & \si{\nano\metre}     \\
                     &             &             & $\Delta\lambda$ & 70          & \si{\nano\metre}     \\
                     &             &             & $n$             & 3           &  --                  \\
                     &             &             & $E_\text{tot}$  & 150         & \si{\joule}          \\
    \bottomrule
  \end{tabular*}
  \caption{Simulation parameters used in the numerical calculations.
    They were chosen as to best match the specifications of planned high-power laser facilities \cite{PAP2013,LEB2013}.}
  \label{tab:comparison.parameters}
\end{table}

\begin{figure*}
  \centering
  \includegraphics{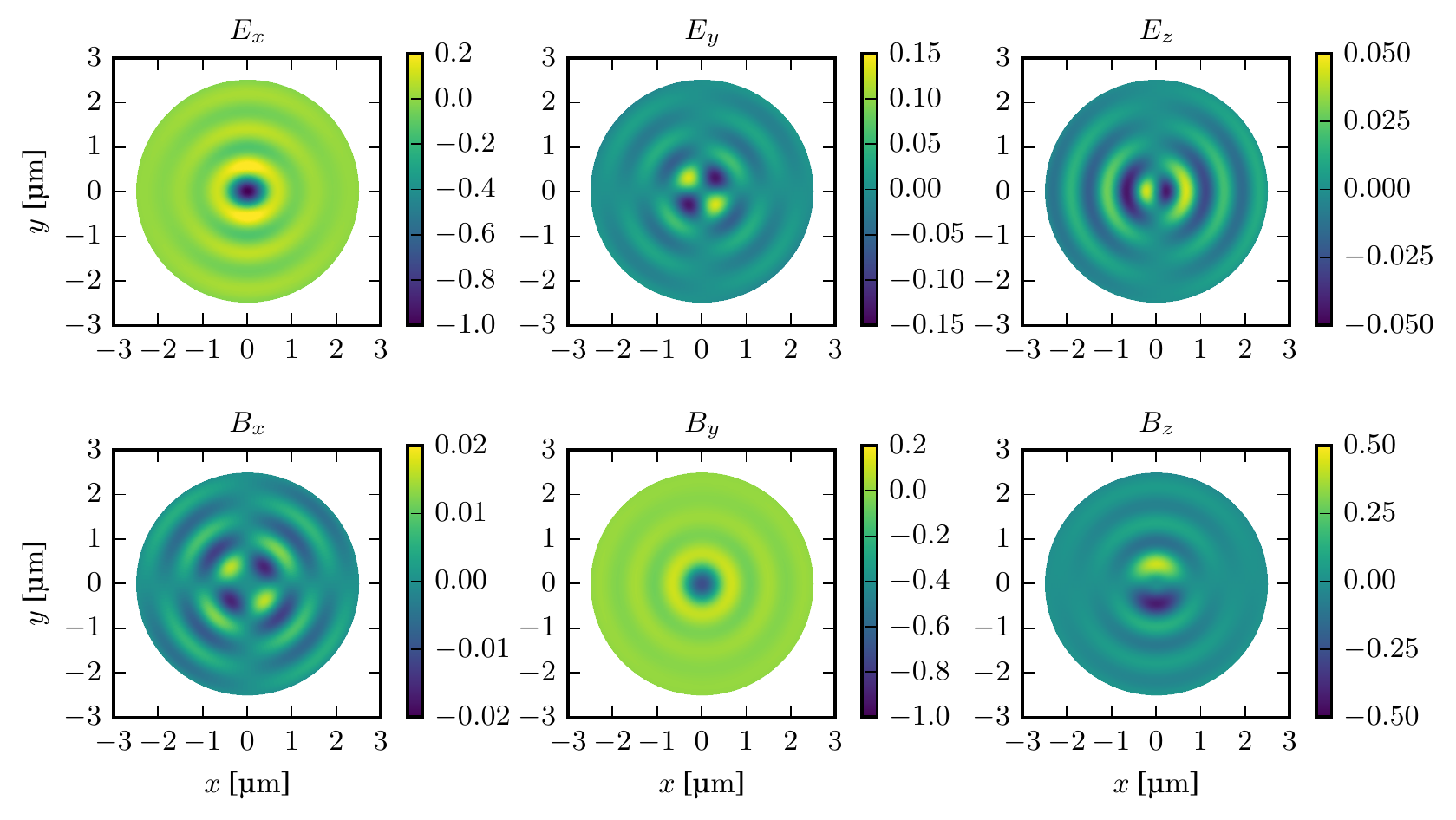}
  \caption{Components of time-dependent electromagnetic field in the focal plane of the
           parabolic mirror at the time at which $I_E$ is maximum.
           The components are scaled with respect to the
           maximum field magnitude across all components. The incident beam
           is polarized in the $x$ direction (horizontal axis in the figure).
           The simulation parameters used to obtain these results are shown
           in Table \ref{tab:comparison.parameters}.}
  \label{fig:comparison.componentsFocalPlane}
\end{figure*}

\begin{figure}
  \centering
  \includegraphics{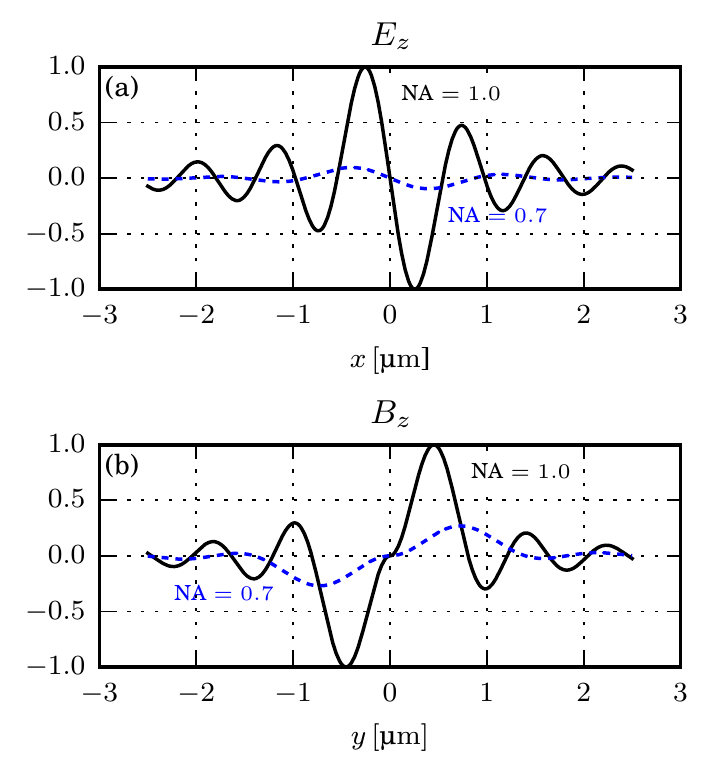}
  \caption{Cuts of the longitudinal components of the electromagnetic field in the focal
  plane for parabolic mirrors of numerical aperture $\text{NA}=1$ and $\text{NA}=0.7$.
  The amplitudes are normalized with respect to the maximum value of the electric and
  magnetic field in the $\text{NA}=1$ case.
  \textbf{(a)} The cut of the electric field is along the $x$ direction.
  \textbf{(b)} The cut of the magnetic field is along the $y$  direction.
  Both electric and magnetic longitudinal field components decrease quadratically
  with numerical aperture (not shown).}
  \label{fig:comparison.longitudinalComponents}
\end{figure}

The structure of the field in the geometric focal plane (Fig.~\ref{fig:comparison.componentsFocalPlane})
exhibits features that cannot be replicated within the paraxial approximation,
nor even by using vacuum solutions of Maxwell's equations, e.g. \cite{SHE1999,Salamin2015}.
The $E_x$ and $B_y$ components, the only non-vanishing components of the incident beam,
dominate the irradiance distribution.
The reflection imparts upon them an elliptical structure with the
major axis in the $x$ direction, i.e. parallel to the laser polarization. This ellipticity
disappears at low enough numerical apertures, i.e. in the paraxial regime.
Moreover, the fields acquire relatively strong longitudinal components, with
the magnitude of $B_z$
being as large as half the $E_x$ component \cite{YOU2000}.
These relatively large longitudinal field components are conspicuously absent from
low numerical aperture simulation results,
confirming that they originate from the strong curvature of the parabolic mirror.
Indeed, comparing transverse cuts of the longitudinal field $E_z$ and $B_z$ in the focal
plane for different numerical apertures ($\text{NA}=1$ and $\text{NA}=0.7$) shows
that the longitudinal fields decrease with lower numerical aperture
(Fig.~\ref{fig:comparison.longitudinalComponents}).
These results highlight the importance of modeling techniques which account for
the reflection off high numerical aperture mirrors, such as the Stratton-Chu
diffraction integrals.

This more accurate beam model could have repercussions in the modeling of physical
processes in the presence of high-intensity laser beams. For instance,
the presence of strong longitudinal components with a smoother beam profile than
the transverse components will affect the trajectories of charged particles interacting
with the laser in the focal spot, due to the different ponderomotive forces
in each direction. Furthermore, the resulting field invariants, which are qualitatively different
than those that occur at low numerical apertures, can also result in the enhancement
of the pair production and vacuum polarization signatures \cite{MON2011}.

The maximum electrical intensity, defined by (in SI units)
  \begin{equation}
    I_E = \frac{1}{2}c\epsilon_0 \bo{E}^2
  \end{equation}
is of interest for SF-QED applications, as most observables strongly depend on this
parameter. It can be shown to scale linearly with the total
energy of the incident beam (via \eqref{eq:theory.normalizationAmplitude}).
For fixed beam waist, simulations show that the intensity increases quadratically with increasing
numerical aperture (not shown).

\begin{figure}
  \centering
  \includegraphics[width=\columnwidth]{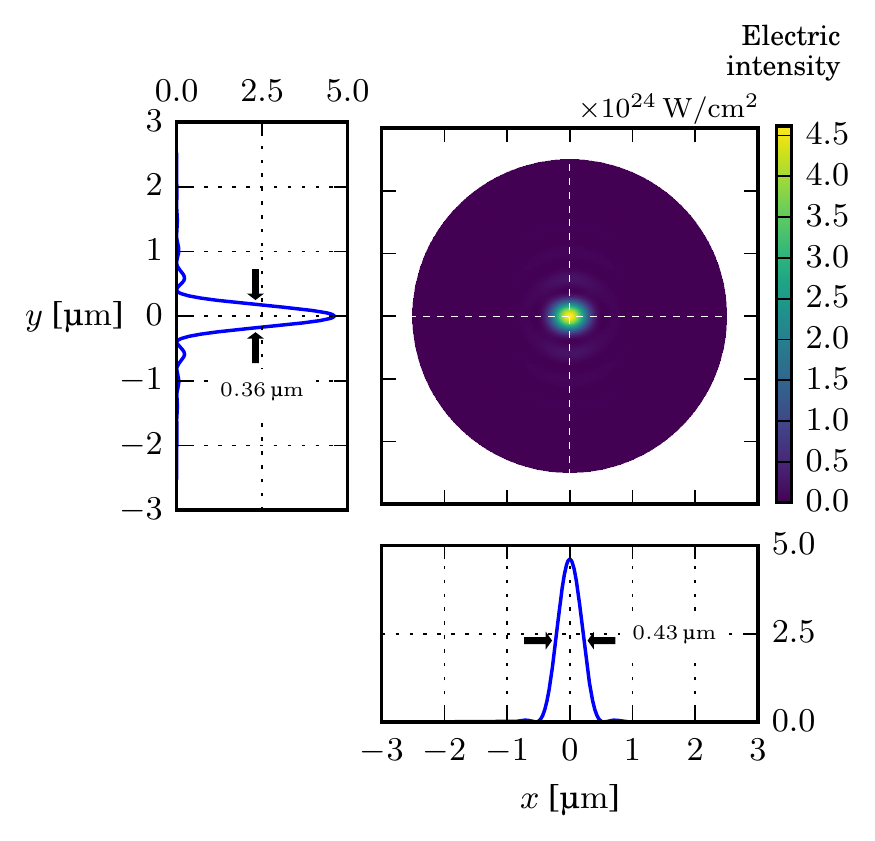}
  \caption{Electric intensity in the focal plane in $\si{\watt\per\cm\squared}$
           and transverse cuts through the focal point (dashed lines). The ellipticity manifests
           itself in different full width at half-maximum in the transverse cuts.
           The resulting eccentricity is approximately $e\simeq0.55$.}
  \label{fig:comparison.electricIntensityFocalPlane}
\end{figure}

In particular, our simulations show that, electrical intensities of up to
$5\times10^{24}\,\si{\watt\per\cm\squared}$ could be obtained
(Fig.~\ref{fig:comparison.electricIntensityFocalPlane}). The electrical intensity has a
discernible elliptical structure. It is caused both by the ellipticity in the
$E_x$ component and, to a smaller extent, by the off-center shape of the $E_z$ component.
At these intensities, it may be feasible to experimentally detect
radiation reaction \cite{BLA2015} and vacuum four-wave mixing \cite{FIL2015a}.

Note that this intensity value does not take into account the possibilitiy
of an imperfect vacuum or of Schwinger pair creation, both of which could
trigger a QED cascade and deplete the laser energy \cite{FED2010,ELK2011}.

\section{Conclusion}\label{sec:conclu}

We used the Stratton-Chu integral representation to model the reflection of temporally short
laser beams off strongly focusing optics. We have shown that while the integral
representation leads to hypersingular integral equations for the reflected
fields in the case of an open surface, the resulting magnetic field integral
equation still has a compact operator and can be solved iteratively with the
Liouville-Neumann series. We have shown that this approach yields the
physical optics approximation. We then generalized this monochromatic technique
to handle temporally short pulses, or, in other words, polychromatic fields.
The formalism is of experimental interest, as its input is the unfocused laser
field coming from the laser system. Usually, this field is experimentally characterized
with a high level of accuracy, and this measurement could in principle be used as
input data for the Stratton-Chu formalism.

We then discussed the development of an efficient parallel, numerical evaluation of the
integrals. It is shown that, unexpectedly, the integrands do not strongly oscillate
in the specific case of a parabolic mirror. This allows the use of simple quadrature
methods (e.g. Gauss-Legendre), instead of specialized quadrature methods.
We have shown that the integral method naturally separates the mirror and focal meshes.
As a consequence, the memory requirements are dramatically lessened as compared to
traditional FDTD, FDFD and FEM methods. This separation also allowed for a highly
efficient parallel implementation of the algorithm. We have shown that our
implementation can be scaled to at least 384 processors.

In the last section, we verified that our implementation converges properly
as a function of the mesh size and that the resulting fields are
solutions of Maxwell's equations in both the frequency and time domain. We also verified
that the algorithm conserves the energy of the incident beam.

We showed that future laser facilities, such as ELI and APOLLON, could obtain focused
intensities on the order of $10^{24}\,\si{\watt\per\cm\squared}$. The dominant
components of the linearly polarized incident beam, $E_x$ and $B_y$, acquire
an elliptical structure upon reflection from a parabolic mirror, with the major axis
being parallel to the polarization. Longitudinal components as large as $0.5E_x$
also appear in the reflected field. Both of these effects are due to the strong
curvature of the high numerical aperture parabolic mirror.

\appendix


\section*{Acknowledgements}

The authors thank S. Payeur, S. Fourmaux, A. Lacha\-pelle and J.-C. Kieffer
for helpful discussions.
J.D. gratefully acknowledges financial support from FRQ\-NT.
Computations were made on the supercomputer MP2 from Université de Sherbrooke,
managed by Calcul Québec and Compute Canada. The operation of this supercomputer
is funded by the Canada Foundation for Innovation (CFI), the ministère de l'Économie,
de la science et de l'innovation du Québec (MESI) and the
Fonds de recherche du Québec - Nature et technologies (FRQNT).
We wish to thank H.~Z. Lu for his tremendous
technical support. We also acknowledge the software packages matplotlib, used
to create some of the figures \cite{HUN2007}, and GNU parallel \cite{Tange2011a}, used
in the data analysis.

\section{Evaluation of the surface integrals at the singular point}
\label{sec:app.limits}

Here, we evaluate the singular part of the integrals of the Stratton-Chu equations,
Eqs.~\eqref{eq:theory.singularLimits}.
The integration is done over a hemispherical surface of radius $R$ centered at
$\bo{r}=\bo{r}'$. We use spherical coordinates $(R,\phi,\theta)$. Note that
$\bo{R}=\bo{r}-\bo{r}'$ in these coordinates.
We then take the limit as $R\rightarrow0$. Since the surface
is small, we suppose that the field does not change on the surface and assume
it takes its value at $\bo{r}=\bo{r}'$ over the whole surface. We can then move it
outside the integral sign. We also ignore the phase of the Green's function, as
it is constant over the hemisphere and goes to 0 after the limit.
The integral over the hemisphere in Eq.~\eqref{eq:theory.regularLimit} becomes
  \begin{align}
    \lim_{R\rightarrow0}\iint_{S_\epsilon}\bo{A}gdS
      &= \lim_{R\rightarrow0}\bo{A}\int_0^{\frac{\pi}{2}}\int_0^{2\pi}\frac{1}{4\pi R} R^2 \sin\theta d\phi d\theta,\nonumber\\
      &= 0.
  \end{align}
The integral is in fact regular at $\bo{r}=\bo{r}'$ because the integration measure
cancels the pole of the Green's function.

In Eq.~\eqref{eq:theory.singularLimit}, the gradient of the Green's function
generates a $R^{-1}$ term and a $R^{-2}$ term. The former does not contribute to the integral
because of the integration measure, while the latter reads
  \begin{align}
    \lim_{R\rightarrow0}\iint_{S_\epsilon}\bo{A}\times\nabla gdS
      &= \lim_{R\rightarrow0}\bo{A}\times\nonumber\\
      &\quad \int_0^{\frac{\pi}{2}}\int_0^{2\pi}\frac{R^2\bou{n}}{4\pi R^2}\sin\theta d\phi d\theta, \nonumber\\
      &= \frac{1}{2}\bo{A}\times\bo{n},
  \end{align}
where $\bo{n}$ comes from the fact that the gradient of the Green's function is
normal to the hemispherical surface.

The integral in Eq.~\eqref{eq:theory.hypersingularLimit} diverges due to the presence
of the double gradient of $g$. However, it is still possible to assign it a finite
value.
To do so, we use the explicit
expression of the double gradient of the Green's function \cite[Eq. (2.61)]{VOL2012}:
  \begin{align}
    \label{eq:singularInt.doubleGradientExp}
    \nabla\nabla g(\bo{r},\bo{r}')  &= -\dyad{I}\left[-\frac{ik}{R}+\frac{1}{R^2}\right]g \nonumber\\
                    &\pushright{+\left(\bou{R}\otimes\bou{R}\right)\left[\frac{3}{R^2}-\frac{3ik}{R}-k^2\right]g}
  \end{align}
where $\dyad{I}$ is the unit dyad, $R=|\bo{r}-\bo{r}'|$, $\bou{R}=(\bou{r}-\bou{r}')/R$
and $\otimes$ is the Kronecker outer product. The contribution of the second line
of Eq.~\eqref{eq:singularInt.doubleGradientExp} to the surface integral
vanishes due to the tensor structure, i.e.
  \begin{align}
    \label{eq:singularInt.dyadVectorProd}
    \bou{R}\otimes\bou{R}\cdot\left(\bou{n}\times\bo{F}\right)&\xrightarrow{r\in S_\epsilon}
      \bou{R}\otimes\bou{R}\cdot\left(\bou{R}\times\bo{F}\right).
  \end{align}
On the hemispherical surface, the dyad $\bou{R}\otimes\bou{R}$ has a single
non-vanishing component in, obviously, the $\bou{R}\otimes\bou{R}$ direction, while the vector
it multiplies, $\bou{R}\times\bo{F}$ only has angular ($\bou{\phi}$ and $\bou{\theta}$)
components. The dyad-vector product
of Eq.~\eqref{eq:singularInt.dyadVectorProd} thus
vanishes. The contribution of the first line of Eq.~\eqref{eq:singularInt.doubleGradientExp}, however,
does not vanish. We expand it in a Laurent-type series and obtain
  \begin{align*}
    \left(-\frac{ik}{R}+\frac{1}{R^2}\right)g
      &\simeq \left(-\frac{ik}{R^2}+\frac{1}{R^3}\right)\left(1+ikR\right) \\
      &= -\frac{ik}{R^2}+\frac{1}{R^3}+\frac{k^2}{R}+\frac{ik}{R^2} \\
      &= \frac{1}{R^3}+\frac{k^2}{R}.
  \end{align*}
Substituting this last expression in the l.h.s of Eq.~\eqref{eq:theory.hypersingularLimit},
we see that the second term vanishes due to the $R^2$ measure and that the
first term formally diverges. To ascribe a finite value to this integral,
we make use of the Hadamard finite part, which essentially drops the diverging term
in a mathematically consistent way \cite{BLA2000}.
Equation (\ref{eq:theory.hypersingularLimit}) thus reads
  \begin{equation}
    \lim_{\bo{r}'\in S} \iint_S \nabla\nabla g \cdot\left(\bou{n}\times\bo{F}\right)dS
     = \HadamardSurf_S \nabla\nabla g\cdot\left(\bou{n}\times\bo{F}\right)dS.
  \end{equation}

\bibliography{stratto}

\end{document}